# Deciphering Charging Status, Absolute Quantum Efficiency, and Absorption Cross Section of Multi-Carrier States in Single Colloidal Quantum Dot

Weiwang Xu[†§], Xiaoqi Hou[‡§], Yongjun Meng[†], Renyang Meng[‡], Zhiyuan Wang[†], Haiyan Qin[‡]*, Xiaogang Peng[‡]*, Xue-Wen Chen[†]*

[†]School of Physics, Huazhong University of Science and Technology, Luoyu Road 1037, Wuhan, 430074, People's Republic of China

[‡]Center for Chemistry of Novel & High-Performance Materials, Department of Chemistry, Zhejiang University, Hangzhou 310027, China

* Corresponding authors:

X.-W. C. (email: xuewen_chen@hust.edu.cn ), H. Q. (hattieqin@zju.edu.cn ) and X. P. (xpeng@zju.edu.cn ).

**ABSTRACT**: Upon photo- or electrical-excitation, colloidal quantum dots (QDs) are often found in multi-carrier states due to multi-photon absorption and photo-charging of the QDs. While many of these multi-carrier states are observed in single-dot spectroscopy, their properties are not well studied due to random charging/discharging, emission intensity intermittency, and uncontrolled surface defects of single QD. Here we report in-situ deciphering the charging status, and precisely assessing the absorption cross section, and determining the absolute emission quantum yield of mono-exciton and biexciton states for neutral, positively-charged, and negatively-charged single core/shell CdSe/CdS QD. We uncover very different photon statistics of the three charge states in single QD and unambiguously identify their charge sign together with the information of their photoluminescence decay dynamics. We then show their distinct photoluminescence saturation behaviors and evaluated the absolute values of absorption cross sections and quantum efficiencies of monoexcitons and biexcitons. We demonstrate that addition of an extra hole or electron in a QD changes not only its emission properties but also varies its absorption cross section.

**KEYWORDS**: colloidal quantum dots, single-dot spectroscopy, exciton, charge states, quantum efficiency, absorption cross section.

Absorption and emission of matter comprise the majority of optical phenomena in our world. Among various kinds of nanoscopic optical emitters, colloidal quantum dots (QDs) have gained great attention due to intense broad-band absorption, tunable narrow-band emission, solution processibility, and compatibility with photonic structures.[1-4] QDs have been demonstrated as building blocks of many functional devices for emerging technologies, such as efficient photovoltaics[5-6], light-emitting diodes,[7-8] lasers,[9-10] and single-photon sources.[11-14] With integration of single QD into plasmonic structures,[15] recent reports showed strong light-QD interactions, such as large enhancement of spontaneous emission[16-18] and vacuum Rabi-splittings.[19] Ultimate success of these applications and demonstrations crucially relies on the absorption or/and emission properties of QDs at single-dot level. In comparison to a single-molecular emitter, density of states of a QD is usually very large and a QD in the charged states can be quite stable.[20-22] Consequently, upon photo- or electrical-excitation, a QD is often found in multi-carrier states.[23-25] Recent studies revealed that these multi-carrier states, especially monoexciton and biexciton of neutral, negatively-charged, and positively-charged QD, might dictate performance of devices, such as in electrically-driven single-photon sources.[26] When extra carrier(s) is (are) acquired for an excited state during excitation and/or emission of a QD, the QD might be in dim emission state(s) due to rapid nonradiative Auger recombination of a three-carrier process.[23,27,28] As one of the possible mechanisms,[29-31] charging may cause a QD to randomly switch among different brightness emission states with constant excitation, i.e., photoluminescence blinking. Despite impressive progress in chemical synthesis of non-blinking QDs for the past years,[32-36] charging, blinking, multiexciton generation and emission still generally occur under relatively intense optical excitation, electric bias or imbalanced current injection.[37-39] These features pose substantial difficulties in quantifying the properties of single QD and become a hurdle for fundamental understanding and rational control of properties of QDs in devices. In the past years, photoluminescence blinking behavior and decay dynamics of single trion, in particular for negative trion, have been studied via controlled electrochemical charging for QDs [28,38,39] or through creating a hole trap for single nanorod with an inorganic nanocrystal.[40] However, for a QD with multiple emitting states under ambient conditions similar to the working environment in real devices, simultaneously identifying the charge sign (if any), absorption cross section, and emission characteristics of the multiple emitting states remains challenging.

There have been several reports in determining the nature of charging for a QD in blinking. For instance, one type of approach is to study the spin dynamics under magnetic field at cryogenic temperatures and vacuum.[41-43] Apart from sophisticated equipment, the measurement conditions of this approach are completely different from those for QDs in practical devices, which may influence the charging behavior. Another approach using electrostatic force microscopy[20-21] requires single QD on a surface for isolation and the implementation of nano-control techniques. Moreover it is unknown whether the charge is around the QD surface or inside the QD, each of which has different consequence on QD's absorption and emission though possesses the same electric response.

To quantify properties of single QD and to understand dynamics of quantum-confined electrons and holes, it is critical to gain precise knowledge of the absorption cross section (ACS)[44-47] and quantum efficiency (QE) of each emitting state, including both charged and neutral emitting states. For QE measurement, a technique based on the Purcell effect[48] has been applied for single molecule[49] and QD at its neutral monoexciton state.[50] Unlike molecules, QDs are more complicated due to charging, blinking, multiexciton and not well-defined transition dipole moments. The technique used in Ref. 50 is limited to neutral QDs with high QE and having out-of-plane dipole moment. The QEs of charged QDs are often obtained by comparing fluorescence intensities with neutral state with the assumptions that the QE of the neutral monoexciton is unity and the ACSs are the same.[42,50,51] In fact, recent reports show that the monoexciton QEs of some non-blinking QDs have large heterogeneities, ranging from 20% to near unity.[18,52]

Here we report in-situ spectroscopic studies at ambient conditions to identify and decipher the neutral, negatively charged, and positively charged states and simultaneously determine their absolute ACSs and QEs. We first uncover very different photon statistics of three types of charging states in single QD and unambiguously identify their charge signs. We then demonstrate distinct PL saturation behaviors of the three charge states. Moreover, by calibrating the detection efficiency of our setup with single terrylene molecule[53] (well-known workhorse in single-molecule spectroscopy since 1990s), we manage to obtain charge-dependent absolute ACSs and QEs of monoexciton and biexciton.

To set the ground for discussion, we consider a simplified model of band-edge emission of neutral, negatively-charged, and positively-charged states in a QD. Upon excitation by a short laser pulse with energy $J$ and spot size $A$, the QD generates electron-hole pairs according to the Poisson distribution and the average number is given by $\langle N \rangle = \sigma_i J / \hbar \omega A$, where $\sigma_i$ is the ACS at excitation photon energy $\hbar \omega$ and $i$ (o, - or +) refers to a specific state (neutral, negatively-charged, or positively-charged state). The generated multiexcitons transit to lower-order states via radiative decay or nonradiative decay dominated by the Auger recombination. We denote the Auger decay rates for negative and positive trions as $\kappa_A^-$ and $\kappa_A^+$, respectively. By counting the trion recombination pathways, we obtain the Auger rates of the $m$-th multiexciton for neutral, negatively-charged and positively-charged states[24] as $m^2(m-1)(\kappa_A^- + \kappa_A^+)/2$, $m(m+$ $1)(m\kappa_A^- + (m-1)\kappa_A^+)/2$ and $m(m+1)((m-1)\kappa_A^- + m\kappa_A^+)/2$, respectively. For radiative recombination of band-edge emission at room temperature, we ignore the fine-structure effects and consider for electron two conduction-band states with S = 1/2 spin degeneracy and for hole four valence-band states with mixed spin and orbital angular momentum degeneracy.[54,55] Thus the radiative decay rates of the *m*-th (*m*>1) multiexciton for neutral, negatively-charged and positively-charged states are (c.f., Supporting Information of Ref. 55) $min(m,4)2\kappa_{x,r}$, $min(m,4)2\kappa_{x,r}$ and $min(m+1,4)2\kappa_{x,r}$, respectively, where $\kappa_{x,r}$ is the radiative decay rate of neutral monoexciton. Note that the positively-charged states for *m*<4 have more radiative recombination channels. The QEs drop dramatically as the order of multiexciton increases and therefore the emissions from the three states will all saturate as the excitation increases. The ratios of the QEs of biexction to monexciton for neutral, negatively-charged and positively charged states read

$$\frac{Q_{bx}^o}{Q_x^o} = \frac{4(\kappa_{x,r}+\kappa_{x,nr})}{4\kappa_{x,r}+\kappa_{x,nr}+2(\kappa_A^-+\kappa_A^+)} \quad (1)$$

$$\frac{Q_{bx}^-}{Q_x^-} = \frac{4\kappa_{x,r}+2\kappa_{x,nr}+2\kappa_A^-}{4\kappa_{x,r}+\kappa_{x,nr}+6\kappa_A^-+3\kappa_A^+} \quad (2)$$

$$\frac{Q_{bx}^+}{Q_x^+} = \frac{6\kappa_{x,r}+3\kappa_{x,nr}+3\kappa_A^+}{6\kappa_{x,r}+\kappa_{x,nr}+6\kappa_A^++3\kappa_A^-} \quad (3)$$

where $\kappa_{x,nr}$ is the intrinsic nonradiative decay rate while $Q_x^i$ and $Q_{bx}^i$ are the QEs of monoexciton and biexciton, respectively. From Eq.(2) and (3), one notes that the sign of charge has a direct influence on $Q_{bx}^i/Q_x^i$, which can be used together with PL decay curves to identify the sign. $Q_{bx}^i/Q_x^i$ is directly linked to the second-order photon correlation function $g^{(2)}(\tau)$ at low excitation through the area ratio of the central peak to the side peak in pulsed excitation.[55] For $(\kappa_A^-, \kappa_A^+) \gg (\kappa_{x,r}, \kappa_{x,nr})$, one has $Q_{bx}^-/Q_x^- = 1/(3+3\kappa_A^+/2\kappa_A^-)$ and $Q_{bx}^+/Q_x^+ = 1/(2+\kappa_A^-/\kappa_A^+)$. For QDs with symmetric conduction and valence bands or QDs with tighter confinement for holes having $\kappa_A^+ \geq \kappa_A^-$ (e.g. standard CdSe and CdSe/CdS QDs[23,56]), a rule of thumb for the charge sign is that: negative if the $g^{(2)}(\tau)$ area ratio is less than 0.22 and positive if more than 0.33. It should be pointed out that we applied a statistical model for treating the Auger recombination rates of neutral biexciton, charged biexciton, and multiexcitons, which is an approximation. Ideally, one should apply more sophisticated models that consider various elementary points, such as state symmetry, spin structure, Coulomb interactions, the shape and nature of confinement potential. [4,30,56-59] Although the simplification induced error in the Auger rate may grow as the order of multiexciton increase, the contribution of the high-order multiexciton emission to total photon count rate is minor due to the small quantum efficiency a well as the excitation levels employed in this work. Here we focus on the monoexciton and biexciton states of neutral and charged states, whose Auger rates can be quite well described by the simple model as discussed later in the measurements on PL decays and photon statistics (see also Section 1 and 2 of Supporting Information). The influence due to the simplified treatment of the high-order multiexciton recombinations on the extraction of physical parameters such as QE and ACS should be limited to a very small percentage.

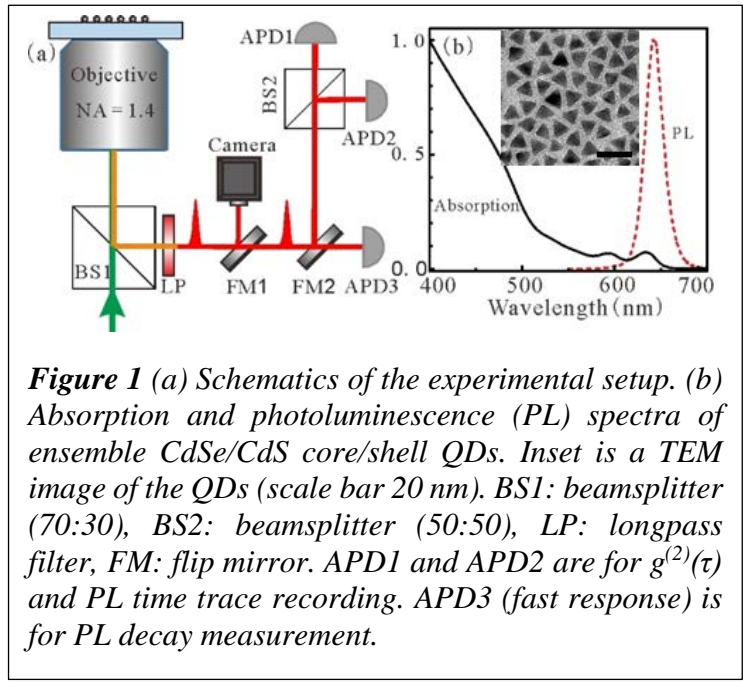

*Figure 1 (a) Schematics of the experimental setup. (b) Absorption and photoluminescence (PL) spectra of ensemble CdSe/CdS core/shell QDs. Inset is a TEM image of the QDs (scale bar 20 nm). BS1: beamsplitter (70:30), BS2: beamsplitter (50:50), LP: longpass filter, FM: flip mirror. APD1 and APD2 are for $g^{(2)}(\tau)$ and PL time trace recording. APD3 (fast response) is for PL decay measurement.*

Our single-dot spectroscopy experiments were conducted with diluted phase-pure zinc-blende core/shell CdSe/CdS QDs (4.2 nm core and 9 layer shells)[29,60] on a coverslip under a home-built inverted microscope setup as illustrated in Figure 1a. The absorption and emission spectra of the ensemble are shown in Figure 1b, where the inset displays the TEM image of the QDs. The QE of the QDs in solution from ensemble measurements by using an integrating sphere system was close to 98%. The microscope (NA=1.4) together with various detectors and time-correlated single-photon counting (TCSPC) unit provided access to a wide range of optical measurements, including PL time trajectory, PL decay dynamics, $g^{(2)}(\tau)$, back focal plane imaging (BFP), and saturation measurements. In all measurements, we used a pulsed laser (~30 ps, 532 nm wavelength) with a repetition rate of 2.1 MHz and an almost diffraction-limited spot to address single QDs (c.f., Section 6 of Supporting Information).

Figure 2a shows a fraction of a typical PL time trace from a single QD under low excitation (photons absorbed per pulse $<N>$ ~ 0.05). The intensity unit is in kcps (kilo counts per second). The QD is non-blinking as confirmed by the histogram in Figure 2b obtained from the whole time trajectory. At higher excitation, probability of photo (dis)charging increased[37] and the QD switched randomly among neutral and charged states as depicted in Figure 2c. The histogram plots (Figure 2d and 2e) for the time trace in Figure 2c reveal that there are three emissive states involved. Among them, one should be the neutral bright state

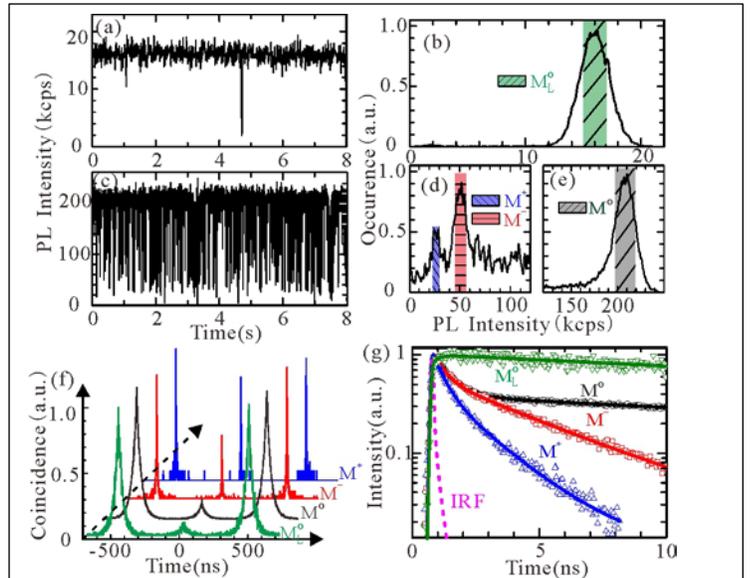

*Figure 2 (a) PL time trace from a single QD under low excitation ($<N>$ ~ 0.05) and (c) higher excitation ($<N>$ ~ 1.0). (b) Histogram of the PL intensity for the time traces at low excitation and (d)-(e) at higher excitation (f) $g^{(2)}(\tau)$ curves at low excitation ($M_L^o$), and for neutral state ($M^o$) and two charged states ($M^-$ and $M^+$) at higher excitation. (g) PL decay curves of instrument response function (IRF) and the corresponding states in (f). Solid lines are the fits to the measurements.*

(denoted as $M^o$) and the other two should be the charged states (at the moment denoted as $M^-$ and $M^+$). The emission was recorded by two APDs and registered to the TCSPC unit. We deciphered the emission according to the average counts of $M^i$ and used the photon-detection events indicated by the shaded bands to obtain $g^{(2)}(\tau)$ for each state, which are illustrated by the color-coded traces in Figure 2f. The trace $M_L^o$ at $<N>\sim0.05$ is also plotted in Figure 2f. The ratios of the areas under the central peak to the side peak are 0.087, 0.115, 0.250 and 0.496 for $M_L^o$, $M^o$, $M^-$ and $M^+$, respectively. For the neutral state, this ratio increased by about 32% from $<N>\sim0.05$ to 1.0. Due to non-blinking at low excitation, we cannot obtain $g^{(2)}(\tau)$ for the charged states. It is known that the area ratio increases with the excitation almost linearly for $<N>$ smaller than 1.0 and the slope decreases as the ratio of the QEs of biexciton to monoexciton increases.[55] In addition, as discussed later, the ACSs of the charged states are smaller by 8%-15%. We estimate the area ratios at low excitation to be around 0.20 and 0.45 for $M^-$ and $M^+$, respectively.

To unambiguously identify the sign of charge, we recorded the PL time trajectory with a fast APD, deciphered the emission states and separately obtained their PL decay curves as in Figure 2g (c.f., Section 1 of Supporting Information). The instrument response function (IRF) is also shown. The PL decay curves are fitted by convoluting a three-exponential decay model including the IRF. The long components revealed by the fitting should be monoexciton lifetimes of $M^o$ (32.3 ns), $M^-$ (3.99 ns), and $M^+$ (1.92 ns), which are very robust against the fitting errors. From $g^{(2)}(\tau)$ measurement at low excitation and monoexciton lifetime of $M^o$, one can obtain the Auger rate of neutral biexciton around 1.31 ns$^{-1}$. This value is very close to the Auger rate of neutral biexction, i.e. two times of the sum of the Auger rates of $M^-$ monoexciton (0.19 ns$^{-1}$) and $M^+$ monoexciton (0.46 ns$^{-1}$) (c.f., Section 2 of Supporting Information). One can rule out the possibility that $M^+$ state has the same sign but more charges as compared to $M^-$ state, because otherwise the Auger rate of $M^+$ should be at least three times of the value of M⁻ monoexciton, contradictory to the measured value. We then re-examine the estimated area ratios of the $g^{(2)}(\tau)$ curves for the charged states at low excitation and compared them with theoretical values predicted by Eq.(2-3). The above analysis leads us to identify that M⁻ state is negatively charged while M⁺ is positively charged.

We next aimed to determine absolute ACSs and QEs of monoexcitons and biexcitons for M⁰, M⁻ and M⁺ states. We first evaluated the total detection efficiency of our setup with single terrylene molecules.[49,53,61] The sample was a thin (~ 25 nm) para-terphenyl crystalline film doped with a low concentration of terrylene molecules, which have nearly vertical electric dipole (VED) orientations.[62] The quantum efficiency of terrylene in para-terphenyl was measured to be near unity[49] and its emitting state (i.e., ON-state) is considered as purely radiative.[61] By counting the photons at the ON-state, we obtained fluorescence saturation curves for single molecules at pulsed excitation with a fixed repetition rate. By fitting the saturation curves for five individual molecules (c.f., Section 3 of Supporting Information), we could determine the detection efficiency for single molecule to be 16.3%±0.2%. For single QD, we then corrected the obtained efficiency by considering the emitting dipole orientations, the para-terphenyl thin film thickness for molecules, the emission spectrum of the QDs, the transmission of the optics, and QE of the APD. The QDs typically have degenerate two-dimensional dipole orientations, which can be separated into contributions from horizontal electric dipoles (HEDs) and VEDs. The fluorescence at the BFP of the objective lens was projected onto a camera and shown as an inset of Figure 3a. The cross section along a dashed line in the BFP image can be translated into angular emission distribution[13,61] as plotted in Figure 3a, where the simulated distributions of a HED and VED are also shown. We obtained the portions of HED and VED for the QD by fitting the measured angular emission distribution and then evaluated the collection efficiency by the objective. By considering the spectral dependence of the optics and detectors, we determined the total detection efficiency of the setup for QDs (Section 4 of Supporting Information).

We characterized the emission of neutral, negatively-charged and positively-charged states in single QD. Our QDs showed excellent photostability[29] and the excitation-power dependent emission had good reproducibility.[37] The emission traces of single QD at each excitation power were recorded and analyzed with suitable time bins to distinguish the three states. In particular, to decipher $M^+$ and $M^-$ states, we applied the following guidelines to choose the time bin: (i) the bin size should be much shorter than the average duration of the emitting state; (ii) photon counts per bin for the least bright emitting state should be well

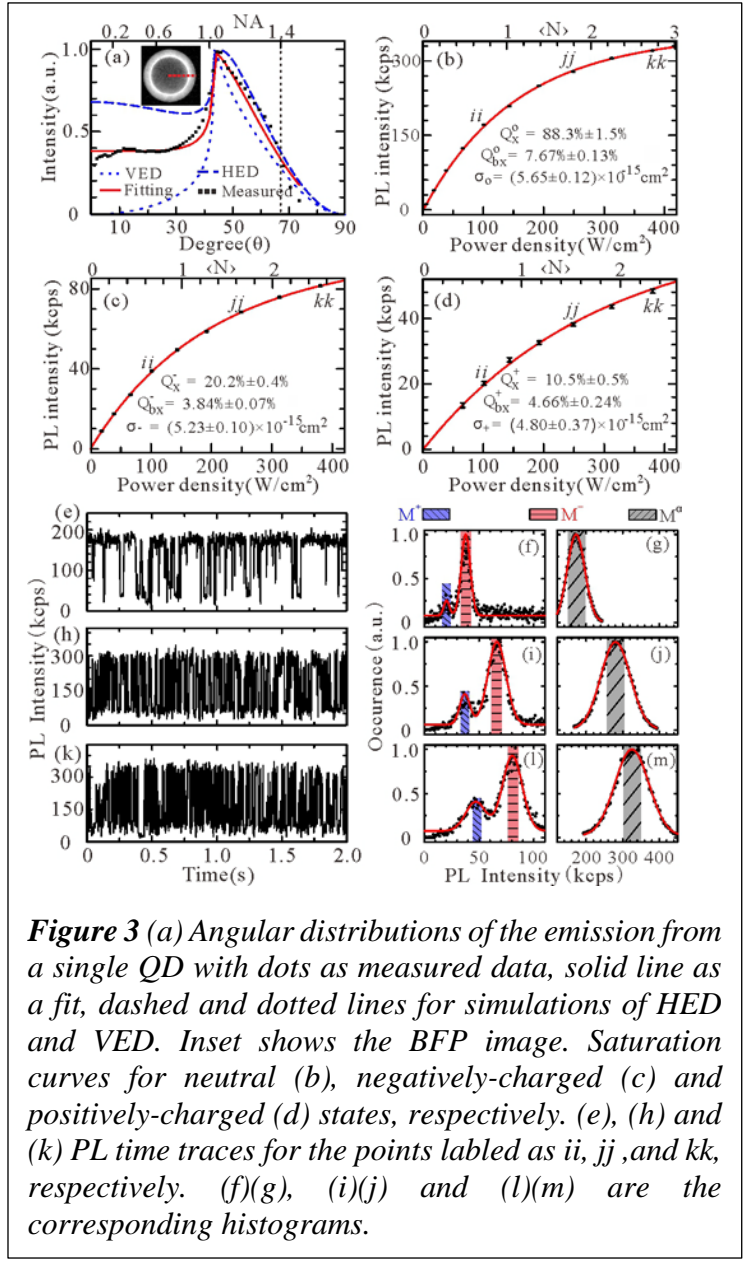

*Figure 3 (a) Angular distributions of the emission from a single QD with dots as measured data, solid line as a fit, dashed and dotted lines for simulations of HED and VED. Inset shows the BFP image. Saturation curves for neutral (b), negatively-charged (c) and positively-charged (d) states, respectively. (e), (h) and (k) PL time traces for the points labled as ii, jj ,and kk, respectively. (f)(g), (i)(j) and (l)(m) are the corresponding histograms.*

above the shot noise level. As the state duration decreases and photon count rate increases along the increase of the excitation, the time bin should be adjusted accordingly instead of being constant.

To be specific, we have chosen the time bin to give an average count of ~ 50 per bin for M$^+$ state at each excitation level. This means a time bin in the range of 1.0 ms to 3.0 ms for the whole series of measurements. As explained in detail in Section 5 of Supporting Information, such time bins followed the above two guidelines and allowed us to decipher and quantify different emitting states.

Figure 3b-3d show the measured count rates as a function of excitation power density for a representative dot at M$^o$, M$^-$ and M$^+$ states, respectively. As examples, for data points labeled as *ii*, *jj* and *kk*, we show their PL time traces in Figure 3e, 3h and 3k, respectively. The background counts were less than 0.7 kcps for the largest excitation used in the experiment. The relatively high detection efficiency of our setup for single QD enabled us to measure the three saturation curves within 7 minutes, which minimized the measurement uncertainty due to issues like the drift of focus. The histograms depict the normalized occurrence as a function of the photon counts for the PL time trace on the same row. For M$^o$, we only collected the bins as effective events for the histogram with at least 10 consecutive bins that have photon counts larger than 60% of average counts. One clearly sees that as the excitation power increases the three emission states labeled with shaded rectangles move towards right (brighter emission). The emission peaks of three different states in histogram plots were identified by fitting with one/two Gaussian functions as indicated by the red traces (Section 5 of Supporting Information). The detected photon counts with error bars shown in Figure 3b-3d for each excitation power were associated with the peaks and uncertainties of the states identified in the histogram plots. We fit the data in Figure 3b-3d based on the theory described above. The numerical models and fitting procedures are given in Section 6 of Supporting Information. The fitting parameters are the monoexciton QE and ACS. While for M$^o$, $Q_{bx}^o/Q_x^o$ was obtained from the area ratio in g$^{(2)}$($\tau$) curve measured at low excitation, the area ratio in g$^{(2)}$($\tau$) for M$^-$ and M$^+$ states at low excitation were evaluated through $Q_{bx}^o/Q_x^o$ and the monoexciton lifetimes of M°, M⁻ and M⁺ states (Section 1 of Supporting Information), which are robust. The solid traces in Figure 3b-3d are the fitted saturation curves for M°, M⁻ and M⁺ states, respectively. The absolute QEs of monoexciton and biexcitons and ACS are displayed in the corresponding figures for each state and the errors are mainly due to the uncertainties from the fitting and the separation for the charged states at high excitations. The QE of M° state in air from single-dot measurement is slightly lower than the ensemble value, which is understandable since the air environment might alter surface states or/and nonradiative channels of the QD.[60] With similar radiative rate but faster Auger rate, positive trion possesses a QE lower than the negative trion[23,56]. The faster Auger rate for positive trion is probably due to higher density of states for the hole and more confined hole wavefunction, which leads to larger Coulomb interactions.

Due to distinct values of $Q_{bx}^i/Q_x^i$, one clearly observes that the "curvatures" of the saturation traces are different so that the neutral state is the first to show saturation while M⁺ state exhibits nonlinear effect at much higher excitation power. The values of ACS for M° state agree reasonably with simple calculations.[44] We also estimated the ACS for M° state from ensemble measurement of QDs in solution.[45] The ACSs from ensemble and single-dot measurements are within the same order of magnitude. Recent experimental results revealed that the ACS from the ensemble measurement seemed to be significantly underestimated for zinc-blende CdSe QDs.[46] Moreover, at the excitation wavelength of 532nm, the ACSs of M°, M⁻ and M⁺ states are not the same but with an additional charge the ACS is decreased. The M⁺ state seems to have the smallest ACS possibly due to the tighter confinement for the holes in CdSe/CdS core/shell QDs where the addition of a hole has a relatively larger effect on the absorption. An extra charge in a QD and the missing charge trapped on the QD surface may have static stark effect on the energy levels.[63] We believe the decline of ACS for M⁻ and M⁺ should be mainly attributed to the bleaching of exciton transition caused by the presence of an extra electron in the conduction band or an extra hole in the valence band.[64,65] Other factors such as band shift caused by Stark effect should be less crucial to influence the ACS of the QDs. In the present study, since the excitation light at 532 nm close to the transition from the valence band to the $1P_e$ state rather than to the $1S_e$ state, the absorption bleaching at this wavelength should be much less than that at the band edge (50%). However, the exact degrees of bleaching for $M^+$ and $M^-$ states are hard to estimate.

We did measurements on a set of nine single QD and found that all of them showed three distinct $M^o$, $M^-$ and $M^+$ states. We then saturated the three states for each dot and determined the absolute values of ACSs, the monoexciton QEs, and the ratio for the QEs of biexciton to monoexciton. The results are summarized in Figure 4. One observes that the charge-dependencies of the ACS are the same for all the dots although the amounts of change are different from dot to dot. For monoexciton QEs, the neutral-state values have less fluctuation from dot to dot in contrary to the charged states. Among the three states, $M^+$ state has the lowest monoexciton QE. The largest monoexciton efficiency of $M^-$ measured in the series is about 20%. The ratios of QEs of biexciton to monoexciton for the three states are displayed in Figure 4c. One sees that the ratio increases sequentially from neutral, negatively-charged, to positively-charged states for the dots as discussed in the theory section.

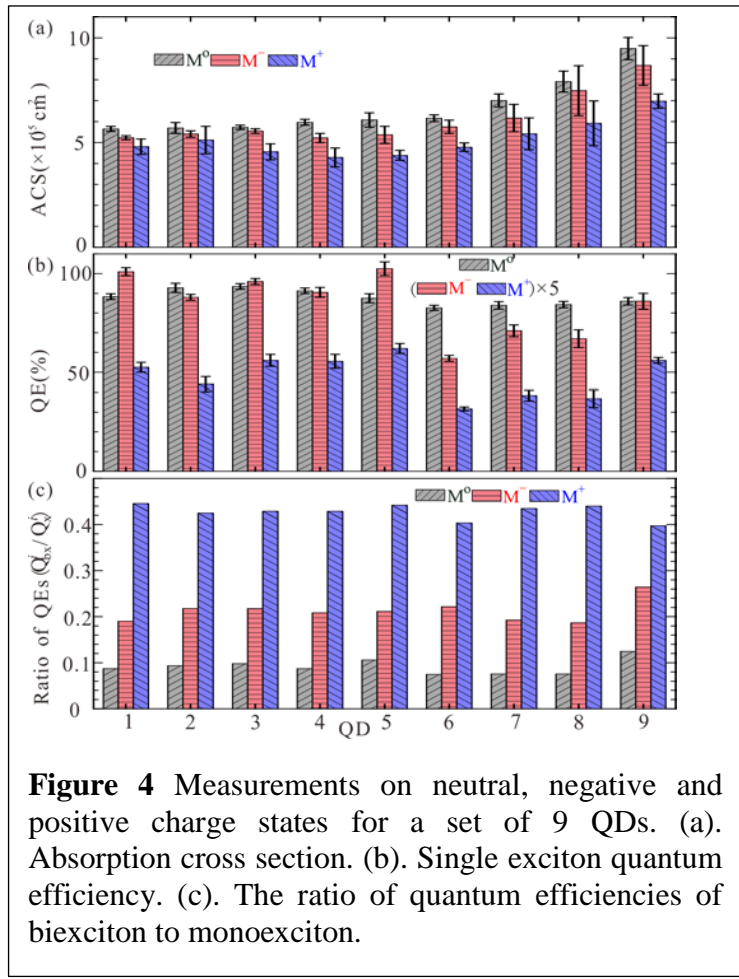

**Figure 4** Measurements on neutral, negative and positive charge states for a set of 9 QDs. (a). Absorption cross section. (b). Single exciton quantum efficiency. (c). The ratio of quantum efficiencies of biexciton to monoexciton.

In summary, we have carried out single-dot spectroscopic studies on neutral, negatively-charged, and positively-charged states of single QD to identify the sign of the charge, absorption cross section, and emission properties of each type of charged states. We showed distinct fluorescence photon statistics and saturation curves for the three types of charged states. By calibrating the detection efficiency with single molecule, we obtained charge-sign dependent absolute ACSs and QEs of the monoexcitons and biexcitons of all three states. Our experimental results demonstrate how an additional charge in a QD substantially influences the absorption and emission properties of single QD. Our method is a simple in-situ spectroscopic technique and can be applied to study single QD at ambient conditions as like in operating devices. We believe the results shall enable better understanding of the photophysics of QDs at various working regimes and allow accurate evaluations of strategies in enhancing single and biexciton emissions of single QD[4,16,18,66] and Auger recombination engineering.[4,56-59]

**ASSOCIATED CONTENT**

***Supporting Information**: Additional information regarding the analysis of Auger rates, ratio of quantum efficiencies of biexciton to monoexciton, the calibration of the system detection efficiencies, choice of time bin for resolving the two charged states, the state-dependent PL saturation models and the error analysis. The Supporting Information is available free of charge on the ACS Publications website.

AUTHOR INFORMATION

**Corresponding Authors**

*Email: xuewen_chen@hust.edu.cn (X.-W. C)

*Email: hattieqin@zju.edu.cn (H.Q.)

*Email: xpeng@zju.edu.cn (X. P.)

**Author Contributions**

§W. Xu and X. Hou contributed equally to this work.

**Notes**

The authors declare no competing financial interest.

**ACKNOWLEDGMENT**

We acknowledge financial support from the National Natural Science Foundation of China (11474114, 21573194), the Thousand-Young Talent Program of China, the National Key Research and Development Program of China (2016YFB0401600), and Huazhong University of Science and Technology.

**TOC Figure**

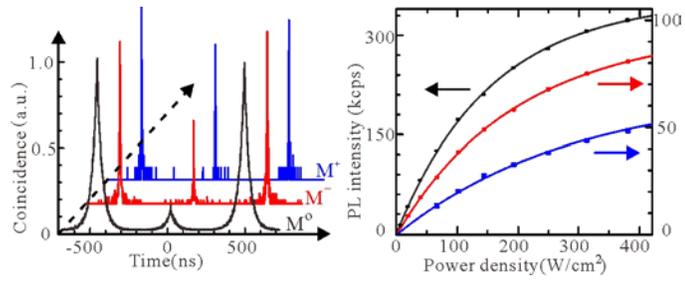